**Title**: Chauhan Weighted Trajectory Analysis of combined efficacy and safety outcomes for risk-benefit analysis

**Running Head**: CWTA of combined outcomes for risk-benefit analysis

**Authors**:

1. Utkarsh Chauhan, MD – Department of Medicine, University of Calgary; Calgary, Alberta, Canada; uchauhan@ualberta.ca

2. Daylen Mackey, BSc – CW Trial Analytics; Edmonton, Alberta, Canada; daylen.j.mackey@gmail.com

3. John R Mackey*, MD FRCPC, Faculty of Medicine and Dentistry, University of Alberta; Edmonton, Alberta, Canada; jmackey@ualberta.ca

*Corresponding author

**Abstract**

**Purpose:** Analyzing and effectively communicating the efficacy and toxicity of treatment is the fundamental basis of risk benefit analysis (RBA). There is a need for more efficient and objective tools. We apply Chauhan Weighted Trajectory Analysis (CWTA) to perform RBA with superior objectivity, power, and ease of communication.

**Methods:** We used CWTA to perform 1000-fold simulations of RCTs using ordinal endpoints that captured both treatment efficacy and treatment toxicity. RCTs were stochastically generated with 1:1 allocation at defined sample sizes and hazard ratios. We first studied the simplest case simulation of 3 levels each of toxicity and efficacy (a 3 x 3 matrix). We then simulated the general case of the advanced cancer trial, with efficacy graded by five RECIST 1.1 health statuses and toxicity graded by the six-point CTCAE scale to create a 6 x 5 matrix. Finally, the 6 x 5 matrix model was applied to a real-world dose escalation phase I trial in advanced cancer.

**Results**:  Simulations in both the 3 x 3 simplest case matrix and the 6 x 5 advanced cancer matrix confirmed our hypothesis that drugs with both superior efficacy and toxicity profiles synergize for greater statistical power with CWTA RBA than either signal alone. The CWTA RBA 6 x 5 matrix meaningfully reduced sample size requirements over CWTA efficacy-only analysis. Despite a small sample size, application of the matrix to each of the seven cohorts of the dose finding phase I clinical trial provided objective and statistically significant validation for the dose subjectively selected by the trialists.

**Conclusion**: CWTA RBA, by incorporating both drug efficacy and the trajectory of drug toxicity, provides a single test statistic and summary plot that analyzes, visualizes, and effectively communicates the risk-benefit assessment of a clinical trial. CWTA RBA requires fewer patients than CWTA efficacy-only analysis when the experimental drug is both more effective and less toxic. Our results show CWTA RBA has the potential to aid the objective and efficient assessment of new therapies throughout the drug development pathway. Furthermore, its distinct advantages over competing tests in visualizing and communicating risk-benefit will assist regulatory review, clinical adoption, and understanding of therapeutic risks and benefits by clinicians and patients alike.

**Introduction**

***Risk-benefit analysis in therapeutic assessment***

Risk-benefit analysis (RBA) is the fundamental evaluation of a therapeutic intervention. In broad terms, RBA addresses the questions "Is this intervention more likely to help than harm? Is intervention A better than intervention B, when both efficacy and safety are considered?"

RBA is essential to the design, ethical review and regulatory approval to start a study, where the potential benefits to patients are predicted to potentially outweigh the risks [1]. Subsequently, during the conduct of clinical trials, repeated RBAs are performed, often by involved investigators in early phase trials, or by independent data monitoring committees in phase II, III, and registrational studies, to ensure that the potential benefits of a clinical trial continue to outweigh the risks to participants. On completion of a trial, the risk-benefit assessment is the key metric on which the decision to advance through additional later stage trials, or seek regulatory approval, is based. Finally, the decisions of individual clinicians, insurance payers, and clinical guideline committees are predicated on the careful assessment of risk-benefit for individual patients and populations. In most cases, the RBA is a subjective assessment by experts informed of the totality of the safety and efficacy data, without the use of a formal statistical analytic tool that incorporates both safety and efficacy signals [2].

In the context of new drug development, RBAs are hampered by several practical problems. In early-stage drug trials (first-in-human interventions, first-in-class drugs, and dose-finding studies), the nature, severity, and duration of adverse events is largely unknown, as pre-clinical models have only limited predictive ability for clinical toxicities [3,4]. Furthermore, patient benefit in early-stage trials is generally not expected until effective doses are established, so formal risk-benefit calculations in this setting are difficult to pre-specify. Typically, the aggregated experience of the patients in these studies is assessed subjectively, with particular emphasis on safety signals rather than efficacy. Finally, sample sizes are typically small and not amenable to standard statistical analytic techniques.

Even after key drug toxicities are identified and the recommended phase II doses are established, ongoing monitoring of risk-benefit still presents challenges in phase II, III, and phase IV settings. RBA is either formally or informally conducted by iterative reviews of patient outcomes during the clinical trial, typically presented as tables of most common and most severe adverse events, summary tables of efficacy signals, and lengthy and complex individual patient narratives [1]. These safety and efficacy data are complex to communicate, lack a unified statistical methodology that incorporates both, and are not readily displayed visually. In consequence, it is difficult to quickly grasp the timing, the reversibility, and the duration of toxicities from the standard presentation of clinical trial results while balancing them with the timing and magnitude of clinical efficacy experienced by patients on trial therapy. Furthermore, RBA is particularly complicated in the setting of a disease where the natural history of the illness varies in severity over time, independent of treatment effects, where the aggregated assessment of risk and benefit is more difficult than either assessment in isolation.

***Limitations of Kaplan-Meier analysis for efficacy and toxicity assessments***

The Kaplan-Meier analysis with logrank testing [5] is the gold-standard method for analysis of efficacy outcomes in most trials, including those in cancer patients.  The KM estimator has many advantages: it is a nonparametric method, making it suitable for analyzing time-to-event data that may not follow a normal distribution, and it can handle censored data effectively, where some individuals may not experience the event of interest by the end of the study, ensuring that these individuals are appropriately accounted for in the analysis. The KM estimator is typically used in conjunction with the logrank test, which allows hypothesis testing to compare survival curves between different treatment groups in clinical trials. While the KM estimator provides a robust and widely accepted method for estimating survival functions from lifetime data, it does have several important limitations that reduce its utility. In particular, KM analyzes only a single time-dependent endpoint for each trial participant, and cannot provide a single analysis that incorporates the many key efficacy endpoints that may occur during a cancer patient's clinical course, including disease response, progression, and death. Furthermore, KM analysis can only model unidirectional outcomes (for example progression of disease) and cannot model the bi-directional outcomes such as initial disease response, followed by later disease progression, and eventual death, that represent the typical illness trajectory of patients with advanced cancer on clinical trials [6].

***Limitations of adverse event tables for safety assessments***

The typical output of clinical trial safety assessments are tables of the highest grade of each CTCAE toxicity experienced by study subjects, where the trajectory of these side effects is not plotted for visual inspection, but must be derived from detailed reading of individual narrative reports for affected patients [1,2]. As such, the timing and the rapidity of onset, the reversibility and duration of maximal toxicity, and the rapidity of adverse event resolution is not typically detailed in clinical study reports or clinical trial publications.

***Rationale for CWTA in RBA***

Chauhan Weighted Trajectory Analysis [7] was specifically developed to address the limitations of KM analysis, while retaining a visual and test statistic output that would be easily interpreted by clinical and regulatory communities due to its similarity to KM analysis. A generalization of the KM methodology, CWTA has several advantages over traditional KM analysis. CWTA permits the assessment of outcomes defined by various ordinal grades or stages or clinical severity, which are common in medical settings but challenging to analyze using traditional methods like the KM estimator. Importantly, it facilitates continued analysis following changes in health state which may be bidirectional (disease recovery or exacerbation), providing a more comprehensive view of the trajectory of clinical outcomes [7].

CWTA also retains the merits of KM analysis. It is a non-parametric method with the ability to censor patients who withdraw or are lost to follow-up, ensuring robust analysis of clinical outcomes. CWTA, like KM, provides a graphical summary plot that visually depicts the trajectories of patients over time, making it easy to interpret and compare outcomes between different treatment arms. This is particularly important in the setting of clinical oncology, where oncologists have become accustomed to seeing a single plot for efficacy endpoints of clinical trials, in which the separation of the control and experimental curves conveys meaningful information as to the differential effects of treatment through time. Finally, CWTA introduces a weighted logrank test, a modification of the traditional logrank test, to assess the statistical significance of differences in trajectories between groups, providing a rigorous method for hypothesis testing that is also a familiar analog for clinicians and regulatory agents [7].

Furthermore, CWTA is a more powerful analytic tool than KM analysis. We recently reported that in the setting of advanced cancer efficacy studies, by using all efficacy signals (CR, PR, SD, PD and death), CWTA was markedly more powerful than KM analysis for the standard single endpoints of PFS or OS [8]. Sample size reductions ranged from 15% to 35%, and time to first efficacy signals were reduced by two- to six-fold. Consequently, CWTA is particularly advantageous in deriving early informative efficacy signals in small sample size studies, populations of patients with rare diseases, and studies where enrollment capacity is limited.

We hypothesized that the risk-benefit analyses of many clinical trials are hampered by the lack of a single analytic tool that incorporates both efficacy and safety signals and conveys the aggregated signal in a visually intuitive, statistically tractable, and easily communicated manner. Due to the complexity and subjectivity of the risk-benefit assessment process throughout the drug development process, we sought and developed a tool that can simultaneously incorporate both benefit (efficacy) endpoints and risk (safety) endpoints in a single objective metric, display this visually, and conduct a formal statistical comparison of group outcomes. We developed and modeled this tool in the context of advanced cancer, using standard RECIST 1.1 criteria [9] for efficacy and CTCAE v5.0 criteria [10] for toxicity. For clinical validation with a real-world trial dataset, we then applied the CWTA RBA methodology to an early-stage clinical trial in advanced cancer by incorporating both comprehensive efficacy outcomes, together with detailed daily toxicity outcomes, to evaluate the clinical benefit of a novel therapy.

**Methods**

*Definitions*

To study CWTA risk-benefit analysis, we created simulation models at two levels of complexity.

For the first model, a 3 x 3 matrix outlined in **Table 2**, we defined three efficacy states (healthy, sick, dead) and three safety states (toxicity none, toxicity some, lethal toxicity). The goal of the first model is to test the hypothesis that a drug that both reduces morbidity and has lower toxicity compared to a drug with only one (or neither) of these attributes will demonstrate greater statistical power in CWTA RBA analysis.

Our second model was a 6 x 5 matrix described in **Table 4**. 6 rows correspond to tiers of toxicity based on CTCAE v5.0 [10], which range from no toxicity to fatal toxicity; 5 columns correspond to tiers of efficacy including complete response (CR), partial response (PR), stable disease (SD), and progressive disease (PD), and death defined as per RECIST 1.1 criteria [9]. The goal of the second model is to (i) demonstrate how clinical grading systems can easily be integrated in CWTA RBA and (ii) test the intuition established in the first model that combined efficacy and toxicity analysis offers greater statistical power compared to efficacy analysis alone. The same matrix is later applied to the real world advancer cancer validation study.

*Simulation design*

The simulation studies were generated using Python 3.8 [11]. Study simulations were stochastic processes in which randomly generated numbers are programmed to mirror fluctuating disease response to chemotherapy cycles with weekly measurements of treatment efficacy and toxicity. Patients could only change one efficacy level per month but could transition weekly to any toxicity level; higher likelihoods were assigned for transitions to more proximal scores. Event probabilities were modified between groups as defined by a hazard ratio (HR). We did not model patient dropout. We performed 1000-fold simulations of chemotherapy RCTs at defined sample size (SS) allocated 1:1 to control or intervention and run for a defined number of weeks and used CWTA (weighted logrank test) to determine statistical significance.

*Simplest case simulation – 3 toxicity x 3 efficacy matrix*

We first modeled the simplest possible state for combined efficacy and toxicity assessment, a 3 x 3 matrix with variables defined in **Table 1** and depicted in **Table 2**. We set up simulation scenarios, ran them each 1000 times, and sought statistically significant differences between the arms. Each scenario was run with a sample size of 600 patients for a total of 210 weeks to mirror large scale randomized trial parameters. Hazard ratios of 0.5, 1, and 2 are used for simplicity and reciprocal property.

Simulation of the 3 x 3 matrix for RBAs was performed for each of the five possible scenarios from each of the five possible cases of relative efficacy and toxicity: i) more efficacy, same toxicity, ii) same efficacy, less toxicity, iii) more efficacy, less toxicity, iv) more efficacy, more toxicity, and v) same efficacy, same toxicity. In case i, the experimental drug was more effective than the control drug [HR 0.5 for illness or death from disease progression] but equally toxic [HR 1.0 for illness or death from toxicity]. In case ii, the experimental drug was equally effective as the control drug [HR 1.0] but was less toxic than the control drug [HR 0.5]. In case iii, the experimental drug was both more effective [HR 0.5] and less toxic [HR 0.5] than the control drug. In case iv, the experimental drug was more effective than the control drug [HR 0.5] but more toxic than the control drug [HR 2.0]. In case v, the experimental drug was equally effective and equally toxic when compared to the control drug [HR 1.0 for both].

*Advanced cancer trial simulation: 5 efficacy x 6 toxicity matrix*

This simulation was built on the 6 stages of CTCAE toxicity and the five stages of RECIST 1.1 efficacy outlined in **Table 3**, as per standard methods in advanced cancer trials. This created the matrix shown in **Table 4.** As per our previous publications on using CWTA in cancer efficacy studies, all patients we assigned SD at time 0 and, each month, were capable of response (PR then CR), maintained SD, or irreversible exacerbation to PD or Death [7,8]. We modeled a control group CR rate of ~10% and a PR rate of ~50% to reflect first-line advanced cancer RCTs. Power was evaluated at increments of SS (20 to 320 patients in increments of 30) and HR (0.6, 0.7, and 0.8). The required sample size for i) combined efficacy and toxicity as per **Table 4** and ii) efficacy alone were directly compared.

*Statistical analysis*

CWTA was performed as previously described [7] augmented by a cloud-native cluster, with workloads executed in parallel across optimized Google Compute Engine virtual machines. For each trial, a p-value was computed using CWTA (weighted logrank test). The fraction of tests that were significant (at $\alpha < 0.05$) represents the power of the test (correctly rejecting the null hypothesis that the two groups are the same). At each hazard ratio in our second model, a sample size requirement was interpolated as the threshold to reach 0.8 power, a common standard in RCT design.

*Real world advanced cancer validation study*

Our validation study was an analysis of real-world data from a published dose escalation study in patients with advanced cancer. The PCLX-001-01 study evaluated escalating doses of continuous therapy with an oral anticancer drug, zelenirstat [12]. In this trial, 29 patients were assigned to various dose cohorts as shown in **Table 5**. Patients took zelenirstat until dose limiting toxicity was experienced or until progressive cancer was identified. Progressive cancer was identified by applying RECIST 1.1 criteria to CT scans performed every two months. Adverse events were logged for each day of study participation and defined and graded by CTCAE v5.0 criteria [10]. For this CWTA RBA, we restricted our analysis to symptomatic toxicities that developed on therapy (also known as treatment-emergent toxicities) mapped to CTCAE v5.0. Due to the palliative nature of the population treated in this advanced cancer setting, we excluded asymptomatic laboratory abnormalities. Each day on therapy was assigned a highest-grade symptomatic toxicity as per CTCAE criteria, and health states for each of the 29 subjects were assigned daily for each study day based on the 6 x 5 matrix as per **Table 4**.

Results

*Overview of findings*

We first modeled the simplest possible state for combined efficacy and toxicity assessment, a 3 x 3 matrix. Simulation of the 3 x 3 matrix for CWTA RBAs confirmed all expectations for each of the five possible scenarios mapped to five possible combinations of relative efficacy and toxicity: i) more efficacy, same toxicity, ii) same efficacy, less toxicity, iii) more efficacy, less toxicity, iv) more efficacy, more toxicity, and v) same efficacy, same toxicity. As expected and shown in **Figure 1**, CWTA identified visually different and statistically significantly different (as assessed by weighted logrank testing) curves in cases i through iv; curves were overlapping and the weighted logrank test was non-significant in cases ii, iv, and v. The aggregate results of 1000-fold simulations are shown in **Figure 2**, in which only combination iii, which synergized improved efficacy and lower toxicity, resulted in power beyond 0.8, the typical standard for trial design.

Next, we modeled the case of an advanced cancer trial, in which efficacy signals were categorized as 5 possible ordinal variables as per RECIST 1.1 criteria, and toxicity was classified as per CTCAE 5.0 criteria in which there are six ordinal variables of toxicity. We conducted simulations of the 6 x 5 matrix RBA in **Table 4** across a range of hazard ratios favoring the experimental arm for both efficacy and toxicity (0.6, 0.7, and 0.8) and a range of samples sizes (20 to 320 patients in increments of 30). A single representative trial showing CWTA RBA vs CWTA efficacy only is shown in **Figure 3**, while overall results are shown in **Figure 4**. These results show that RBA outperforms CWTA efficacy only analysis of studies when both efficacy and safety signals are favorable for the experimental drug. Sample sizes required to achieve 80% power were reduced by applying RBA, with larger reductions as hazard ratios approached 1.0; sample size reductions were 17% at a hazard ratio of 0.6 (RBA 44 patients vs. CWTA efficacy 53 patients), 18% at a hazard ratio of 0.7 (RBA 87 patients vs CWTA efficacy 106 patients), and 24% reduction at a hazard ratio of 0.8 (RBA 212 patients vs. CWTA efficacy 279 patients). Sensitivity analysis (data not shown) confirmed that, as expected, improved efficacy can be offset by worsened toxicity in the experimental arm.

Finally, we validated the 6 x 5 matrix CWTA RBA with real world data derived from a completed and published phase I dose escalation clinical trial in advanced cancer [12]. This phase I study involved evaluation of seven doses of a first-in-class small molecule inhibitor of N-myristoylation, zelenirstat**.** In this dose escalation study, 29 patients were administered daily continuous oral zelenirstat at one of seven dose levels **(Table 5).** The trialists then used the standard toxicity assessment methods to define the Maximally Tolerated Dose (MTD) as first cycle Dose Limiting Toxicities (DLTs) occurring at a frequency of 33% or less, after which the efficacy signals were subjectively integrated into the decision to select dose level 6, 210 mg daily continuous therapy, as the recommended phase 2 dose for subsequent studies [12]. We applied the 6 x 5 CWTA RBA matrix to the completed zelenirstat trial database as shown in **Figure 5**. These results confirmed that the recommended phase II dose, 210 mg daily, provided the statistically best combined health status for patients on that cohort, when compared the aggregated health status of the patients treated on other cohorts, and was the only dose level to show significantly better outcomes. Thus, CWTA RBA objectively validated the investigator's subjective choice of cohort 6 as the recommended phase II dose.

**Discussion**

Risk-benefit analysis (RBA) is an essential component to the development and evaluation of any health intervention. It is a complex assessment, representing the integration of efficacy / relative efficacy signals and safety / relative safety signals across the spectrum of clinical trials, regulatory reviews, and knowledge translation to clinical providers. We previously lacked a statistical tool that integrates both safety and efficacy signals, incorporates the trajectory of both disease and toxicity, yields a single test statistic, and readily communicates these outcomes visually.

In this study, we began with simulations of simplified therapeutic trials with three efficacy categories and three safety categories. We demonstrate that this model can produce composite CWTA analyses that accurately reflect the risk-benefit scenarios we modeled, across a range of therapeutic efficacy rates and toxicity rates **(Figures 1, 2).** These results confirmed the intuitive assumptions that CWTA could correctly identify an experimental drug that had better efficacy than the control drug, better toxicity than the control drug, or both better efficacy and toxicity. As expected, more effective but more toxic drugs did not show improved CWTA RBA.

Next, we simulated the more complex scenario of an advanced cancer therapy trial, in which efficacy status was captured with five categories (CR, PR, SD, PD, and Death), which key toxicities were captured with CTCAE v5.0 categorized toxicities ranging from grade 0 (no toxicity) to grade 5 (fatal). In this simulation, efficacy was assessed monthly, while adverse events were modeled with weekly changes, reflecting the real world complexities of an advanced cancer trial. We demonstrated that across a range of clinically relevant efficacy rates, as identified by hazard ratios compared to control therapies, and various assumptions of toxicity rates, CWTA could address these complexities and distill the risk-benefit assessment into a relatively simple single metric and single visualization. Importantly, CWTA RBA was more powerful than CWTA efficacy alone, requiring fewer patients to demonstrate superiority of a drug that is both more effective and less toxic. Given that CWTA efficacy analyses already are substantially more effective than standard Kaplan-Meier analyses of Progression Free Survival and Overall Survival in the settings of advanced cancer [8], this demonstrates that CWTA RBA permits markedly smaller sample sizes than standard KM endpoints in advanced cancer, further reducing subject numbers, time to completion of study, and trial costs.

Finally, we applied CWTA to a real-world clinical trial dataset derived from a dose escalation trial using the first in class agent, zelenirstat. While this drug had not previously been studied in humans, pre-clinical toxicities were primarily those of gastrointestinal events. As such, a pre-specified list of adverse events of special interest (AEIs) and a selection of dose limiting toxicities were assigned prior to launching the study. As the cancer population was treatment refractory, with a median of four prior courses of systemic therapy (ranging up to 8), the population was not healthy and had baseline symptoms including fatigue, neuropathy, and others. Despite the complexities of the trial population and the small number of patients in each cohort, CWTA RBA clearly demonstrated its utility in this complex setting, correctly identifying the go forward dose derived from the standard, subjective analysis of the study. That CWTA achieved statistical significance when the RPD2 cohort (n=7) was compared to non-RPD2 cohort patients (n=22), was a striking validation of the power of CWTA RBA in analysis of trials with small sample sizes.

Our findings are particularly relevant in the context of regulatory efforts to improve dose and schedule optimization in the context of clinical trials. The US FDA's Project Optimus is an initiative by the Oncology Center of Excellence aimed at reforming the dose optimization and dose selection paradigm in oncology drug development [13-15]. Traditionally, dosages for oncology drugs were selected based on the maximum tolerated dose (MTD), which may lead to severe toxicities without additional efficacy benefits. Project Optimus seeks to address this by encouraging early and rigorous dose-finding studies that balance efficacy and safety, minimizing long-term toxicities and improving patient outcomes. CWTA, applied to this effort, would provide statistical rigor and objective criteria on which the optimal dose and schedule of cancer drugs is based.

Effectively communicating the risks and benefits to participants in clinical trials, healthcare professionals, and the public is critical and challenging. Miscommunication can lead to misunderstandings about the safety and efficacy of a drug, impacting patient trust and the overall perception of the drug's value. Currently, communicating clinical benefit and clinical toxicity is difficult, due in part to the lack of methods to integrate both into a single, easily explained metric. For example, the dramatic efficacy of COVID-19 vaccines in reducing hospitalization and death was, in public communications, largely disassociated from the potential for rare, albeit serious, adverse effects [16]. In principle, using CWTA to integrate both efficacy (freedom from hospitalization, ICU, and death) and key safety considerations with serious adverse events (acute vaccine reactions, myocarditis, Guillain Barre syndrome) could provide a single, easily explained, readily visualized difference in health status between a vaccinated and unvaccinated populations.

Our methodology is distinct from decision analytic models, which are mathematical tools used to aid decision-making in situations of uncertainty. Such models help evaluate different decision paths based on their potential outcomes and the probabilities of these outcomes. In particular, Bayesian trials can be structured to assess both safety and efficacy endpoints simultaneously. This is achieved by defining prior probabilities for both the efficacy and safety of the intervention, and then updating these probabilities as data on these endpoints are collected throughout the trial. However, CWTA, unlike Bayesian trials, can give time-dependent toxicity assessments that recover, unlike the typical Bayesian design where toxicities of a certain severity (typically grade 3 or higher) are tallied as discrete isolated incidents, irrespective of their duration. Consequently, CWTA gives a more nuanced and accurate reflection of the patient experience, particularly in settings of transient toxicities which are either self-limited or responsive to therapy. Furthermore, full Bayesian risk-benefit analyses are difficult to communicate and are not reducible to a single test statistic and a simple graphic representation.

In the absence of defined or known key toxicities, such as in the early-stage trial setting, CWTA can be structured to assess efficacy signals in tandem with any toxicities, as we did here with the zelenirstat study, agnostic to the specifics of the toxicity experienced by the study subjects. However, CWTA can be customized to address RBA in the context of phase II trials and beyond, where the key toxicities of the intervention are known prior to study inception. In such a setting, adverse events of special interest could be prospectively incorporated into a study-specific ordinal outcome matrix which ranks the clinical importance of the toxicities, relating them to the clinical importance of efficacy metrics, to generate a combined ordinal ranking that can be used as the primary CWTA RBA endpoint throughout the trial.

Our study has limitations. We specifically addressed the context of advanced cancer trials in our second simulation model and our real-world analysis, in which CTCAE adverse events were integrated into cancer efficacy endpoints. In principle, our findings are expected to be generalizable to other efficacy scales and other toxicity scales, as the improved power of the methodology comes from incorporation of the full dataset in the analyses. We have, however, not explored other matrices which might be more relevant to non-oncology studies. We suspect that the methodology and learnings from this study are generally applicable to medical device trials, public health interventions, and industrial applications, and these will be areas for further research.

Finally, CWTA RBA benefits from very granular data on timing, severity, and resolution of toxicity data; while these data are captured in case report forms, some additional programming is required to extract the trajectory of all adverse events, rather than the usual reporting of rates of individuals with high grade toxicities.

**Conclusion**

CWTA RBA, by incorporating the efficacy and toxicity signals throughout the entirety of the health trajectory of trial participants, provides a single statistic and summary plot that analyzes and effectively communicates the risks and benefits of a clinical trial intervention. CWTA RBA can be fruitfully applied even to small phase I studies, and can provide objective evaluations at any point in a study and across the spectrum of clinical trials of new therapeutics. Our data confirm that using CWTA RBA could aid the assessment of new therapies at multiple levels, including the design phase of clinical trials, interim analyses, adaptive design assessment of alternative therapies, regulatory review, and clinical adoption and communication of therapeutic benefit to clinicians, patients, and funders. We propose CWTA as a flexible and pragmatic tool for aggregate assessment of risk-benefit in the development and comparative assessment of health interventions.

**Acknowledgement:**

We thank Pacylex Pharmaceuticals Inc. for sharing their data on trial PCLX-001-01 / NCT04836195 for the analyses in this paper.

**Tables:**

**Table 1. Ordinal efficacy and safety variables used for the simple 3x3 model.**

| Health Status | Ordinal Value | Type |
|---|---|---|
| Healthy | 0 | Reversible |
| Sick | 1 | Reversible |
| Dead from Disease | 2 | Irreversible and absorptive |
| Nontoxic | 0 | Reversible |
| Toxic | 1 | Reversible |
| Poisoned | 2 | Irreversible and absorptive |

**Table 2. Ordinal efficacy and safety matrix used for the 3 x 3 simplest case simulations.**

|                | Healthy | Sick | Fatal Disease |
|----------------|---------|------|---------------|
| **Nontoxic**       | 0       | 1    | 3             |
| **Toxic**          | 1       | 2    | 3             |
| **Fatal Toxicity** | 3       | 3    | 3             |

**Table 3. Ordinal efficacy variables used for the advanced cancer analysis based on RECIST 1.1 criteria**

| Health Status | Ordinal Value | Type |
|---|---|---|
| Tumor in complete response (CR) | 0 | Reversible |
| Tumor in partial response (PR) | 2 | Reversible |
| Stable disease (at baseline; SD) | 4 | Reversible |
| Progressive disease (PD) | 6 | Non reversible |
| Death (from cancer) | 11 | Non reversible / Absorptive |

**Table 4. Ordinal efficacy and safety matrix used for the advanced cancer analysis.**

All toxicity states are potentially reversible except death.

|          |   | Efficacy |    |    |    |       |
|----------|---|----|----|----|----|-------|
|          |   | CR | PR | SD | PD | Death |
| Toxicity | 0 | 0  | 2  | 4  | 6  | 11    |
|          | 1 | 1  | 3  | 5  | 7  | 11    |
|          | 2 | 2  | 4  | 6  | 8  | 11    |
|          | 3 | 3  | 5  | 7  | 9  | 11    |
|          | 4 | 4  | 6  | 8  | 10 | 11    |
|          | 5 | 11 | 11 | 11 | 11 | 11    |

**Table 5. Dose-escalation Treatment Cohorts in Pacylex zelenirstat phase I dose escalation clinical trial.** Using standard subjective assessment of toxicities and efficacy signals, dose level 6 had been selected by the clinical trialists and published as the recommended dose with which to conduct subsequent clinical trials (reference 6). CWTA provides objective statistical validation for this decision.

| Dose Level | Zelenirstat (mg daily) | Number of subjects | RBA analysis vs aggregated other cohorts | RBA weighted logrank analysis p value |
|---|---|---|---|---|
| 1 | 20 | 3 | equivalent | 0.913 (NS) |
| 2 | 40 | 3 | equivalent | 0.964 (NS) |
| 3 | 70 | 3 | worse | 0.013 (significant) |
| 4 | 100 | 5 | equivalent | 0.406 (NS) |
| 5 | 140 | 3 | equivalent | 0.933 (NS) |
| **6** | **210** | **7** | **better** | **0.034 (significant)** |
| 7 | 280 | 5 | equivalent | 0.553 (NS) |

**Figures:**

**Figure 1. 3x3 simulation of combined efficacy and toxicity assessments.**

Each panel below shows a single representative trial from each of the five possible combinations of relative efficacy and toxicity: i) more efficacy, same toxicity, ii) same efficacy, less toxicity, iii) more efficacy, less toxicity, iv) more efficacy, more toxicity, and v) same efficacy, same toxicity. Only trials i and iii demonstrated statistical significance at α = 0.05.

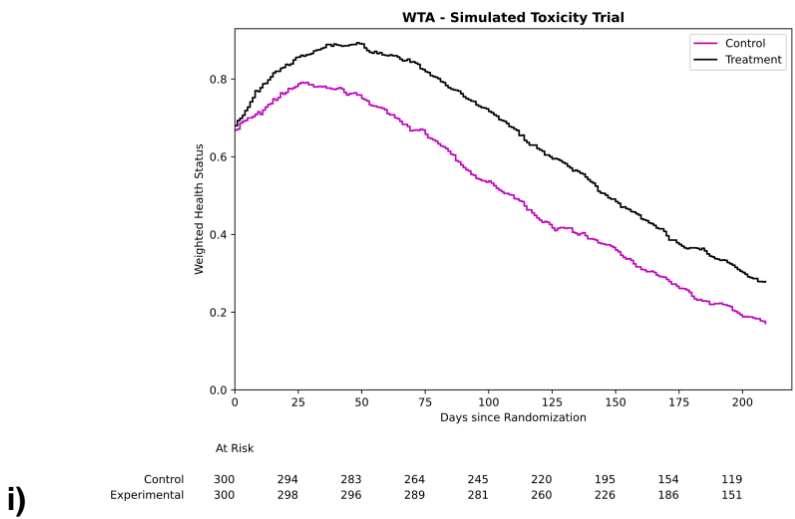

i)

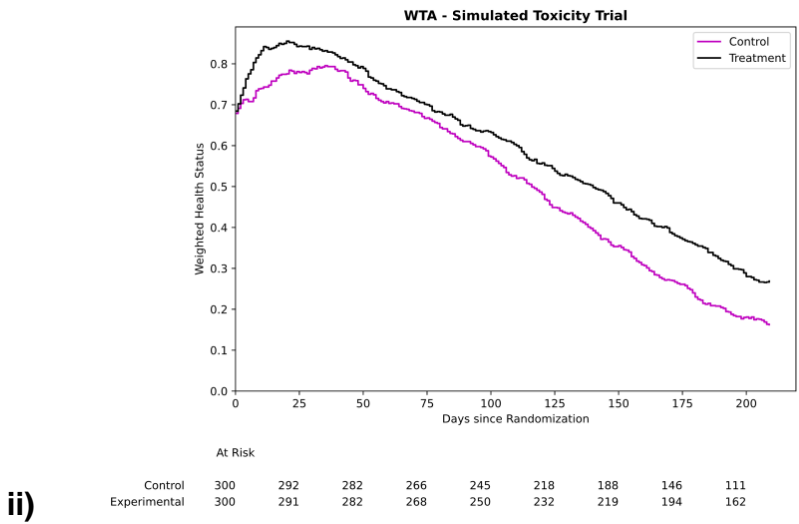

ii)

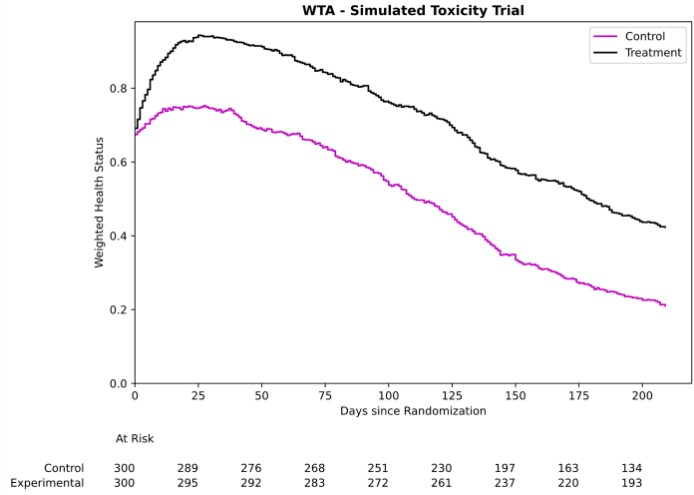

iii)

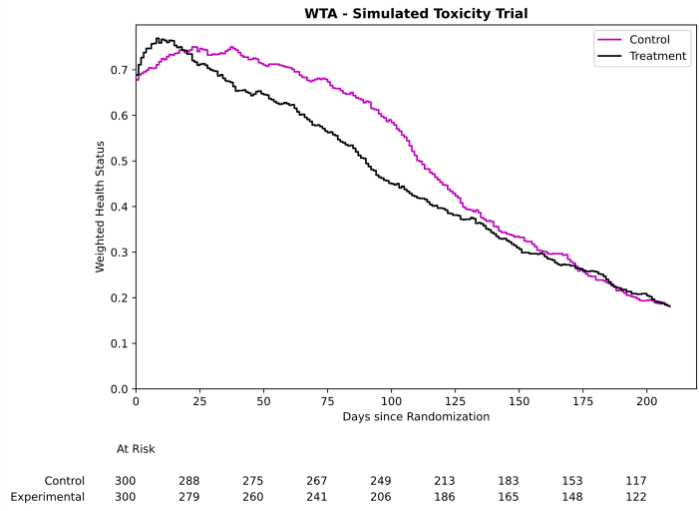

iv)

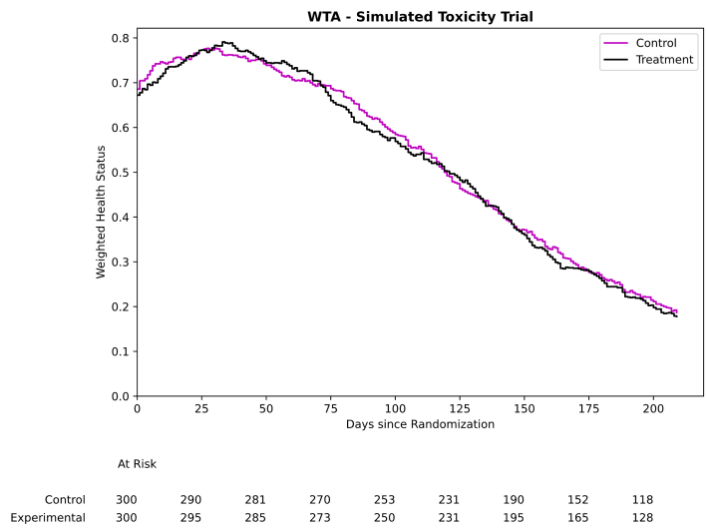

v)

**Figure 2. 3x3 simulation Overall Comparison**

Each horizontal bar represents the overall power of 1000-fold simulated trials defined by five possible combinations of relative efficacy and toxicity: i) more efficacy, same toxicity, ii) same efficacy, less toxicity, iii) more efficacy, less toxicity, iv) more efficacy, more toxicity, and v) same efficacy, same toxicity. Only case iii, where improved efficacy synergized with lower toxicity, resulted in power beyond 0.8 (dashed vertical line), a usual standard for trial design.

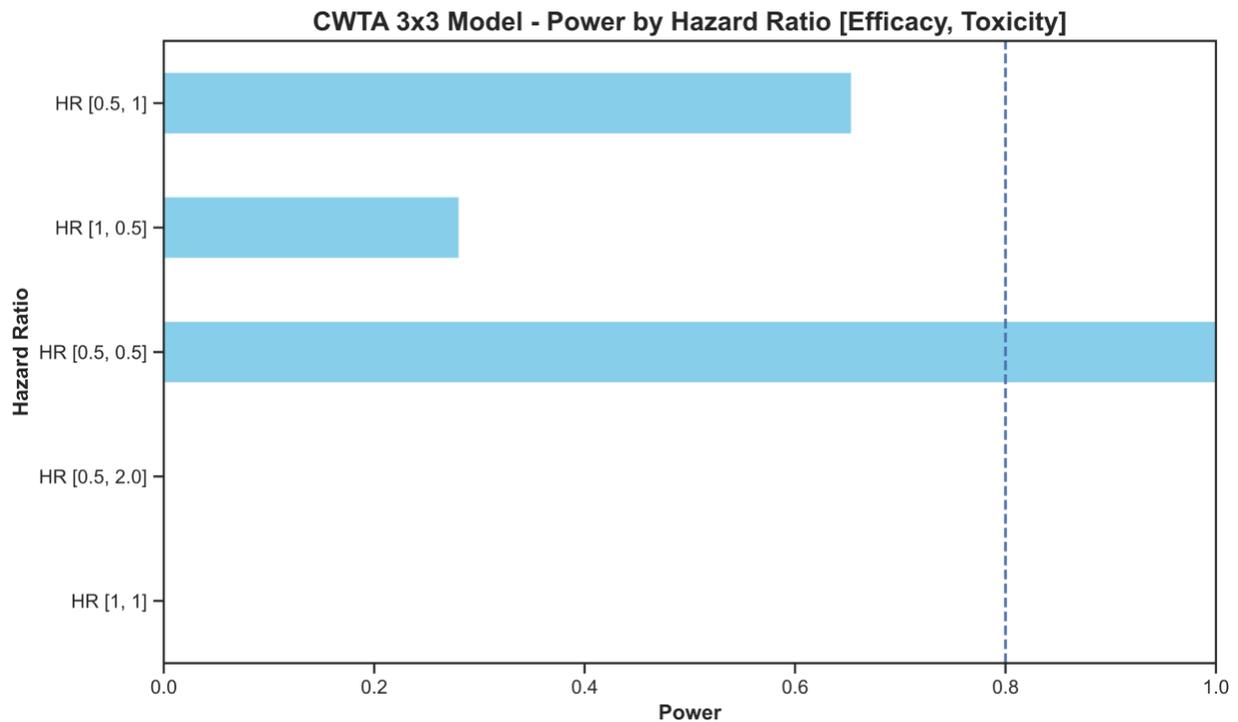

**Figure 3. 6 toxicity x 5 efficacy matrix simulation comparing RBA vs. CWTA efficacy only**

A single representative trial showing RBA vs CWTA efficacy is shown here, where the efficacy HR was 0.8 and the toxicity HR was also at 0.8 for the experimental arm. Here, we see at sample size of 200 subjects, RBA analysis was statistically significant, while the efficacy only CWTA analysis was negative.

i) CWTA Efficacy-Only (p = 0.225)

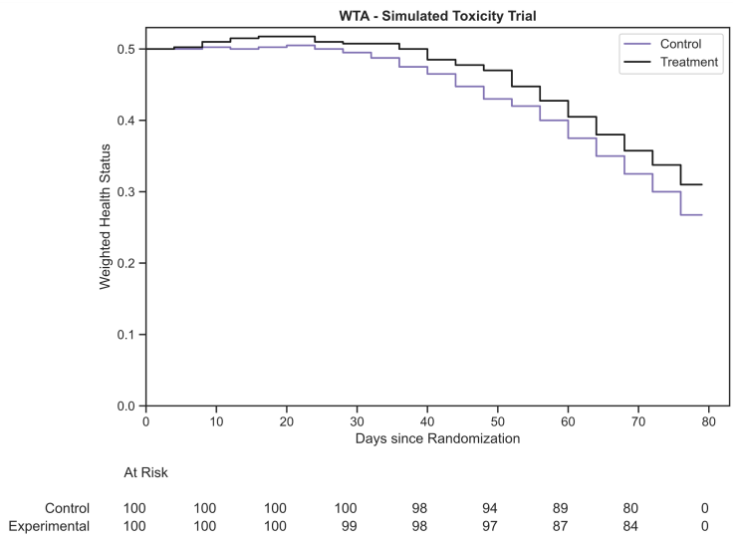

ii) CWTA RBA (p = 0.014)

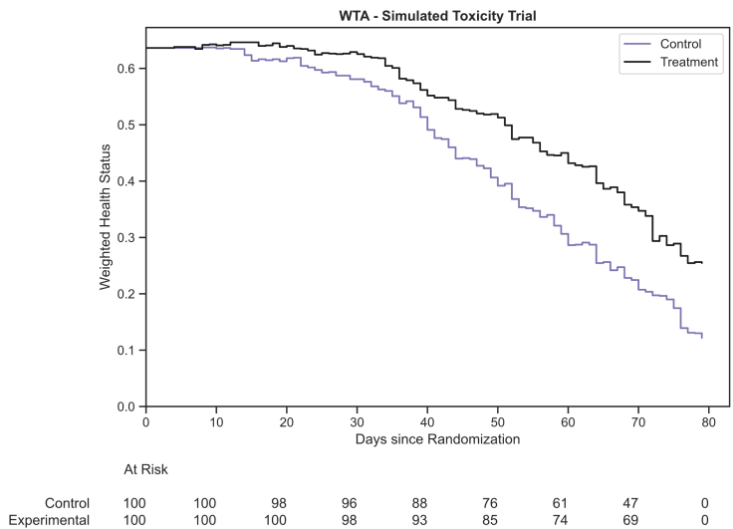

**Figure 4. 6 toxicity x 5 efficacy matrix simulations: power as a function of sample size.** We conducted simulations of the 6 x 5 matrix RBA across a range of hazard ratios favoring the experimental arm for both efficacy and toxicity (0.6, 0.7, and 0.8) and a range of samples sizes (20 to 320 patients in increments of 30) to compare CWTA RBA with CWTA efficacy-alone analysis. Sample size reductions were 17% at a hazard ratio of 0.6 (RBA 44 patients vs. CWTA efficacy 53 patients), 18% at a hazard ratio of 0.7 (RBA 87 patients vs CWTA efficacy 106 patients), and 24% reduction at a hazard ratio of 0.8 (RBA 212 patients vs. CWTA efficacy 279 patients).

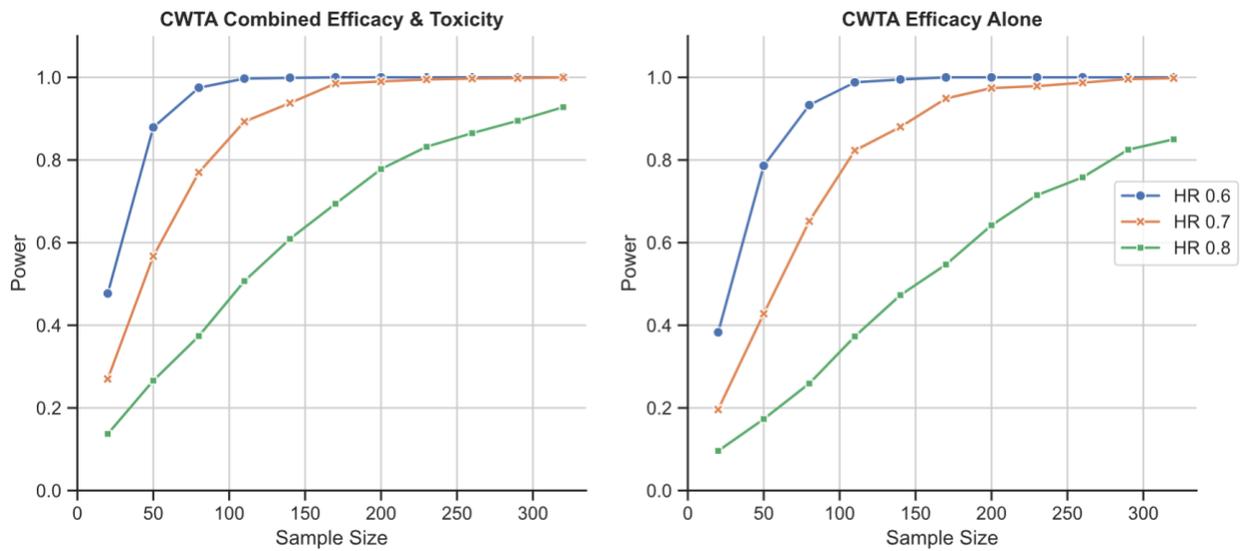

**Figure 5. Zelenirstat dose escalation RBA with 6 toxicity x 5 efficacy matrix analysis.**

Each panel below illustrates the RBA analysis of each cohort vs. the aggregated outcomes of all other cohorts. Patients on cohort 3 did significantly worse than other cohorts, while patients on cohort 6, had significantly better heath status than patients on other cohorts. This result objectively validates the subjective assessment of the clinical trialists, who recommended proceeding with the 210 mg daily dose for subsequent clinical trials.

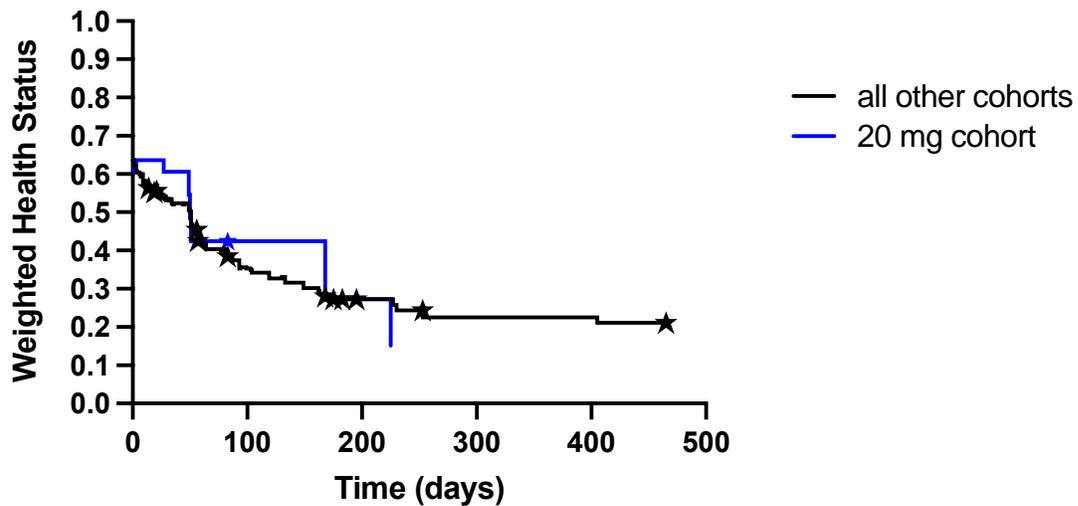

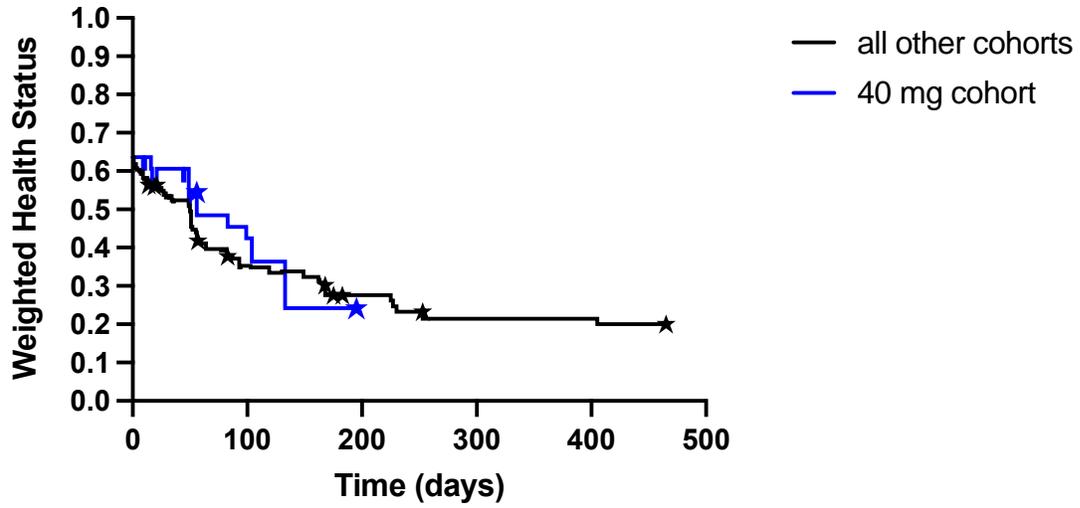

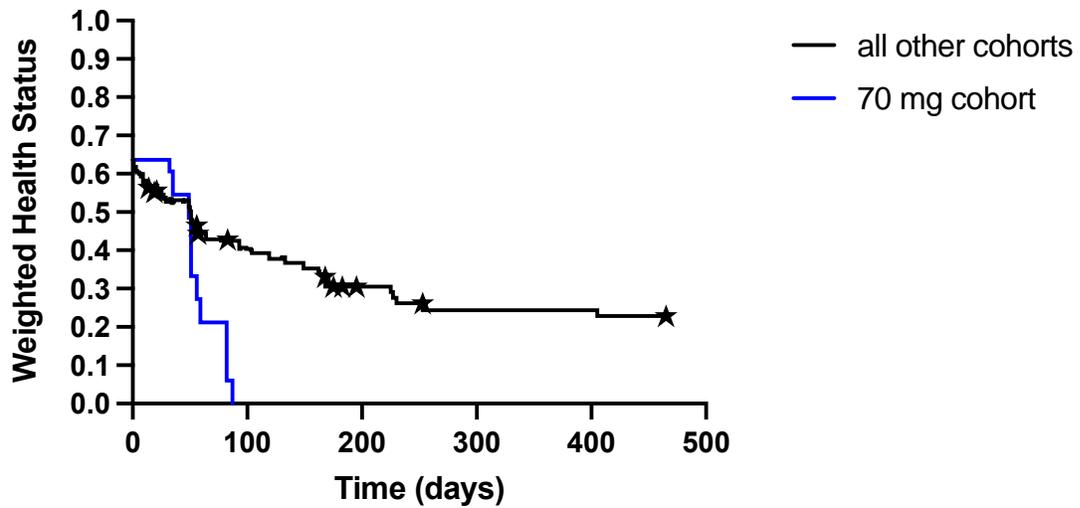

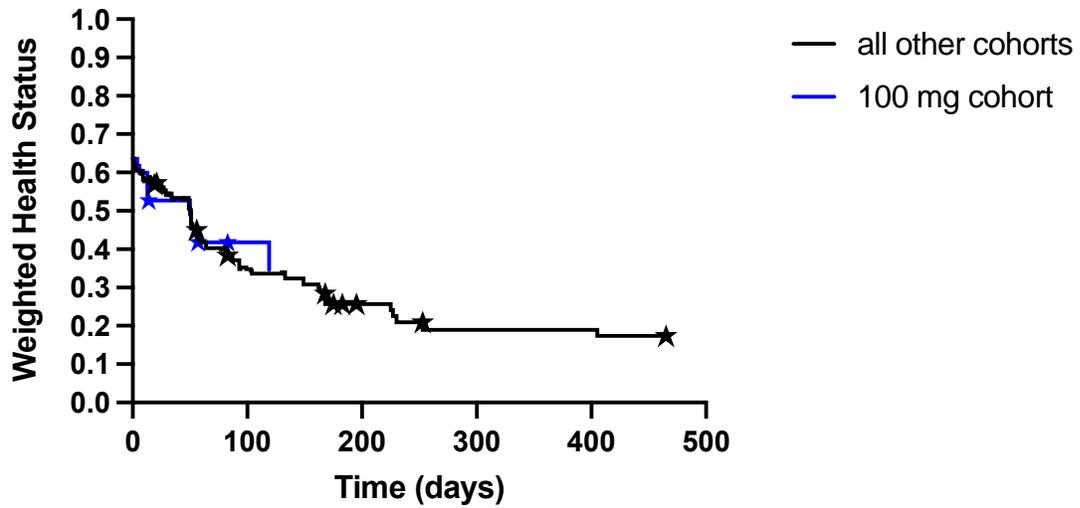

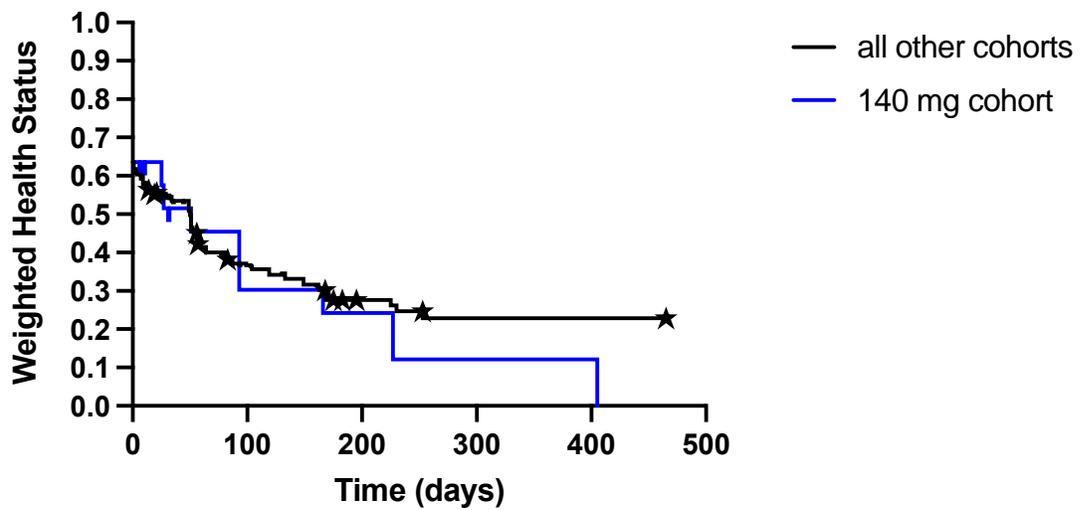

## Weighted Health Status 210 mg vs others

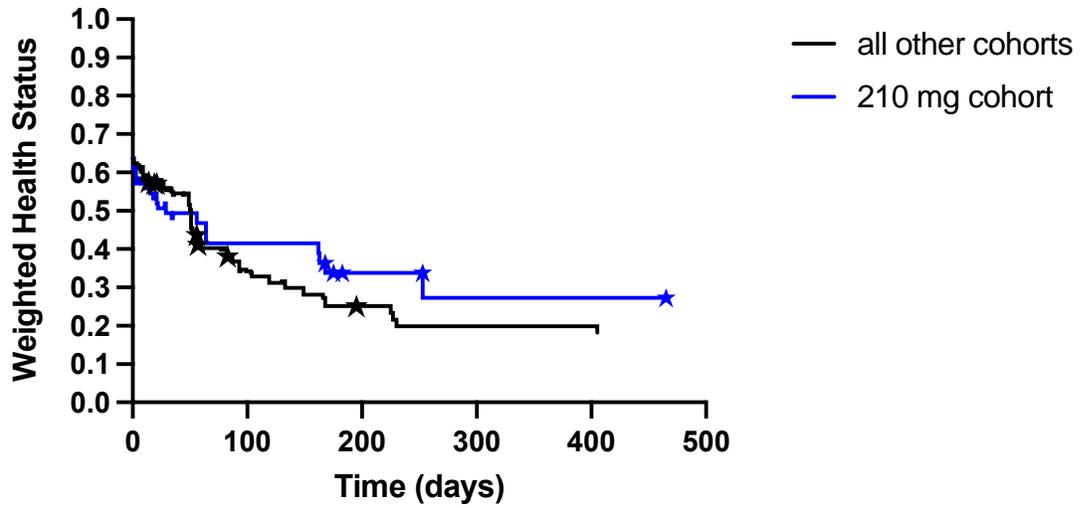

## Weighted Health Status 280 mg vs others

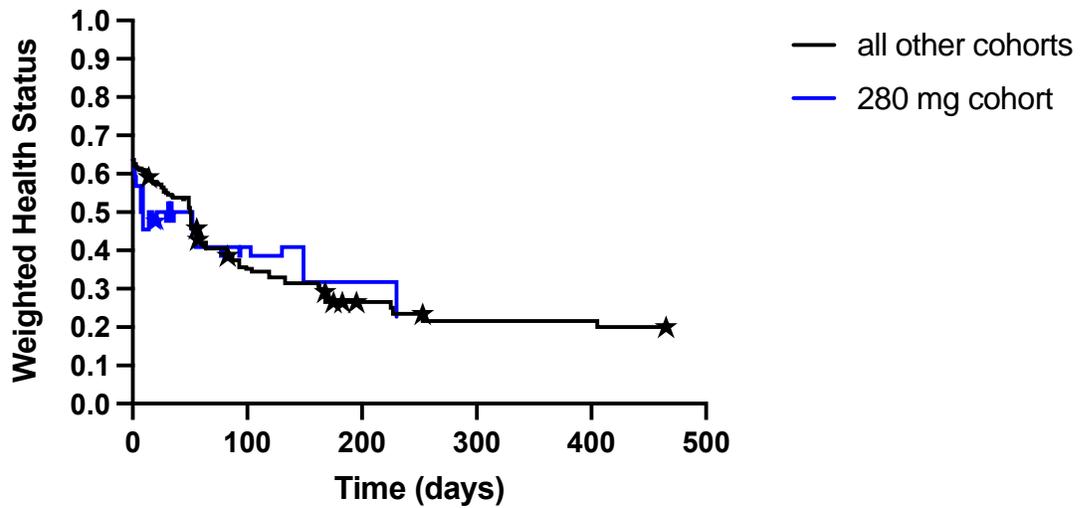